# A GN-model closed-form formula supporting ultra-low fiber loss and short fiber spans

*Mahdi Ranjbar Zefreh[1,2] and Pierluigi Poggiolini[2]*

1 - CISCO Systems, Vimercate (MB), Italy
2 - OptCom, DET, Politecnico di Torino, 10129, Torino, Italy;

## Abstract

We report on a fully closed-form approximation of the GN model for Nyquist WDM systems that extends the range of applicability of previously available formulas to values of fiber loss and span loss that were not previously covered. In particular, so far, a closed-form formula was available for zero loss, and another for span loss of at least about 7 dB. The new formula is accurate over any fiber loss and any span loss value. The interest for this new formula is varied, ranging from hollow-core fibers to improving modeling of standard systems that present lumped loss along installed spans (such as splices or connectors).

## Introduction

Recently, in the quest for new fiber types that may improve on the throughput of current silica-core fibers, new types of fibers have been proposed, such as hollow-core fibers (HCFs, see for instance the nested-antiresonant type [1]). One of their notable features is the potential for achieving in the future values of loss that could be lower than that of silica solid-core fibers. It would then be of interest to assess their performance in hypothetical new systems that took advantage of this and other features of these fibers.

Non-linear interference (NLI) noise modeling, despite the low value of non-linearity in HCFs, would be valuable, to assess their ultimate limitations. To perform such study, a closed-form formula (CFF) for NLI that supports loss values lower than SMF (but non-zero) would be valuable. However, regarding existing CFFs, one is available for zero-loss (Eq.(24) in [2]) and another is available for span loss of at least 7 dB (Eq.(13) in [2]), which may be too high to discuss all the scenarios of potential interest. The 0 to 7 dB range is currently not covered.

A CFF that was able to cover the range 0 to 7 of span loss would also be very valuable in a completely different context. In conventional systems, often times splicing and cabling impairments are present, that cause lumped loss within a single span. One way to deal with the modeling of these impairments is to assume that the span breaks up into multiple sub-spans, where the lumped loss acts as the place where one sub-span ends and another starts [3]. However, since lumped losses may occur anywhere, these sub-spans can be very short and their loss can be significantly lower than 7 dB. So, it would not be possible to model each sub span using the available CFFs. In this particular case, we also argue that the *coherence* of NLI generated in one sub-span with respect to the next, would be high. A proper CFF for dealing with this case should therefore also try to retain the NLI coherence feature.

In this paper we derive new formulas that cover the low (but non-zero) fiber and span loss regime, and also address the case of coherence.

We then thoroughly test the obtained results. The final CFF shows an error below 0.5 dB for any scenario, with respect to the numerically integrated GN model.

## Derivation

We start from GN formula for one single span with an amplifier at the span end which recovers the span loss. The power spectral density (PSD) of nonlinear interference (NLI) is calculated as [2]:

$$G_{NLI}(f) = \frac{16}{27}\gamma^2 \int_{-\infty}^{+\infty}\int_{-\infty}^{+\infty} G_{WDM}(f_1)G_{WDM}(f_2)G_{WDM}(f_1+f_2-f)$$
$$\times \left|\frac{1-e^{(-2\alpha+j\xi)L}}{2\alpha-j\xi}\right|^2 df_1 df_2 \quad \text{Eq. (1)}$$

Where $G_{WDM}(f)$ is the WDM signal PSD which enters the fiber, $\gamma$ is the nonlinear parameter of the fiber, $\alpha$ is the attenuation parameter of the fiber (power evolves as $\exp(-2\alpha z)$ along the fiber), $L$ is the length of the fiber and $\xi$ is:

$$\xi = 4\pi^2\beta_2(f_1-f)(f_2-f) \quad \text{Eq. (2)}$$

Where $\beta_2$ is the second order dispersion parameter of the fiber. For simplicity we consider the whole WDM signal as a rectangular-shaped wide spectrum as:

$$G_{WDM}(f) = \begin{cases} G_{WDM} & -\frac{B_{WDM}}{2} \leq f \leq +\frac{B_{WDM}}{2} \\ 0 & otherwise \end{cases} \quad \text{Eq. (3)}$$

This assumption represents the limit of "Nyquist WDM", that is rectangular individual channel spectra, spaced exactly as the symbol rate.

We then focus on the center frequency of the comb, that is $f=0$ in our relative frequency reference frame, and from Eq. (1) we have:

$$G_{NLI}(0) = \frac{16}{27}\gamma^2 G_{WDM}^2 \int_{-\frac{B_{WDM}}{2}}^{+\frac{B_{WDM}}{2}} \int_{-\frac{B_{WDM}}{2}}^{+\frac{B_{WDM}}{2}} G_{WDM}(f_1+f_2) \times \left|\frac{1-e^{(-2\alpha+j4\pi^2\beta_2 f_1 f_2)L}}{2\alpha - j4\pi^2\beta_2 f_1 f_2}\right|^2 df_1 df_2 \qquad \text{Eq. (4)}$$

Based on the discussion reported in appendix F and fig.24 of [2], to have an analytic expression, we change the 2-D integration area from a lozenge shape (exact) to a square shape (approximate) and Eq. (4) approximately converts to:

$$G_{NLI}(0) \cong \frac{16}{27}\gamma^2 G_{WDM}^3 \int_{-\frac{B_{WDM}}{2}}^{+\frac{B_{WDM}}{2}} \int_{-\frac{B_{WDM}}{2}}^{+\frac{B_{WDM}}{2}} \left|\frac{1-e^{(-2\alpha+j4\pi^2\beta_2 f_1 f_2)L}}{2\alpha - j4\pi^2\beta_2 f_1 f_2}\right|^2 df_1 df_2 \qquad \text{Eq. (5)}$$

When $e^{-2\alpha L} \ll 1$, Eq. (5) is approximately written as:

$$G_{NLI}(0) \cong \frac{16}{27}\gamma^2 G_{WDM}^3 \int_{-\frac{B_{WDM}}{2}}^{+\frac{B_{WDM}}{2}} \int_{-\frac{B_{WDM}}{2}}^{+\frac{B_{WDM}}{2}} \left|\frac{1}{2\alpha - j4\pi^2\beta_2 f_1 f_2}\right|^2 df_1 df_2 \qquad \text{Eq. (6)}$$

The 2-D integral in Eq. (6) is analytically approximated as (see appendices F and G of [2]):

$$\int_{-\frac{B_{WDM}}{2}}^{+\frac{B_{WDM}}{2}} \int_{-\frac{B_{WDM}}{2}}^{+\frac{B_{WDM}}{2}} \left|\frac{1}{2\alpha - j4\pi^2\beta_2 f_1 f_2}\right|^2 df_1 df_2 \cong \frac{\operatorname{asinh}\left(\frac{\pi^2 \beta_2 B_{WDM}^2}{4\alpha}\right)}{4\pi\beta_2\alpha} \qquad \text{Eq. (7)}$$

Therefore having Eq. (7), Eq. (6) becomes:

$$G_{NLI}(0) \cong \frac{4}{27\pi} \times \frac{\gamma^2 G_{WDM}^3}{\alpha\beta_2} \times \operatorname{asinh}\left(\frac{\pi^2 \beta_2 B_{WDM}^2}{4\alpha}\right) \qquad \text{Eq. (8)}$$

## Low Loss Single Span

For notation simplicity we pose:

$$x \triangleq j4\pi^2 \beta_2 f_1 f_2 \qquad \text{Eq. (9)}$$

Also, we define the function:

$$F_1(x) \triangleq \frac{1 - e^{(-2\alpha+x)L}}{2\alpha - x} \quad \text{Eq. (10)}$$

For $|F_1(x)|^2$ as an integrand of 2-D integral in Eq. (5) we do not have an analytic solution. While as we saw from previous section, when the span loss is high, $e^{-2\alpha L} \ll 1$, we will have:

$$F_1(x) \cong F_2(x) \; ; \; \text{for } e^{-2\alpha L} \ll 1 \quad \text{Eq. (11)}$$

where $F_2(x)$ in Eq. (11) is defined as:

$$F_2(x) \triangleq \frac{1}{2\alpha - x} \quad \text{Eq. (12)}$$

And when $|F_1(x)|^2$ is replaced by $|F_2(x)|^2$ in Eq. (6) we can have an closed-form solution.

To have a similar approximate closed-form solution for the low loss case, we consider the following function:

$$F_3(x) \triangleq \frac{A_{eq}}{2\alpha_{eq} - x} \quad \text{Eq. (13)}$$

Where $A_{eq}$ and $\alpha_{eq}$ are two constant parameters with respect to $x$ (also with respect to $f_1$ and $f_2$)

As $F_2(x)$ and $F_3(x)$ have similar forms, Eq. (7) gives us:

$$\int_{-\frac{B_{WDM}}{2}}^{+\frac{B_{WDM}}{2}} \int_{-\frac{B_{WDM}}{2}}^{+\frac{B_{WDM}}{2}} \left|\frac{A_{eq}}{2\alpha_{eq} - x}\right|^2 df_1 df_2$$

$$= A_{eq}^2 \times \int_{-\frac{B_{WDM}}{2}}^{+\frac{B_{WDM}}{2}} \int_{-\frac{B_{WDM}}{2}}^{+\frac{B_{WDM}}{2}} \left|\frac{1}{2\alpha_{eq} - j4\pi^2 \beta_2 f_1 f_2}\right|^2 df_1 df_2$$

$$\cong \frac{A_{eq}^2}{4\pi\beta_2 \alpha_{eq}} \times \text{asinh}\left(\frac{\pi^2 \beta_2 B_{WDM}^2}{4\alpha_{eq}}\right) \quad \text{Eq. (14)}$$

We can write the Taylor series of function $F_1(x)$ (presented in Eq. 10) around $x = 0$ as:

$$F_1(x) = \frac{1 - e^{(-2\alpha+x)L}}{2\alpha - x} = \frac{1 - e^{-2\alpha L}}{2\alpha} + \frac{1 - e^{-2\alpha L} - 2\alpha L e^{-2\alpha L}}{(2\alpha)^2} x + \cdots \quad \text{Eq. (15)}$$

And

$$F_3(x) = \frac{A_{eq}}{2\alpha_{eq} - x} = \frac{A_{eq}}{2\alpha_{eq}} + \frac{A_{eq}}{(2\alpha_{eq})^2} x + \cdots \qquad \text{Eq. (16)}$$

We then choose $A_{eq}$ and $\alpha_{eq}$ in such a way that the two 1st terms of Taylor series of $F_1(x)$ and $F_2(x)$ (presented at Eq. (15) and Eq. (16) respectively) become equal. Therefore:

$$\alpha_{eq} = \frac{\alpha \times (1 - e^{-2\alpha L})}{1 - e^{-2\alpha L} - 2\alpha L e^{-2\alpha L}} \qquad \text{Eq. (17)}$$

$$A_{eq} = \frac{(1 - e^{-2\alpha L})^2}{1 - e^{-2\alpha L} - 2\alpha L e^{-2\alpha L}} \qquad \text{Eq. (18)}$$

Therefore, approximately replacing $F_1(x)$ with $F_3(x)$ in Eq. (5):

$$\begin{aligned} G_{NLI}(0) &\cong \frac{16}{27}\gamma^2 G_{WDM}^3 \int_{-\frac{B_{WDM}}{2}}^{+\frac{B_{WDM}}{2}} \int_{-\frac{B_{WDM}}{2}}^{+\frac{B_{WDM}}{2}} \left| \frac{1 - e^{(-2\alpha + j4\pi^2\beta_2 f_1 f_2)L}}{2\alpha - j4\pi^2\beta_2 f_1 f_2} \right|^2 df_1 df_2 \\ &\cong \frac{16}{27}\gamma^2 G_{WDM}^3 \int_{-\frac{B_{WDM}}{2}}^{+\frac{B_{WDM}}{2}} \int_{-\frac{B_{WDM}}{2}}^{+\frac{B_{WDM}}{2}} \left| \frac{A_{eq}}{2\alpha_{eq} - j4\pi^2\beta_2 f_1 f_2} \right|^2 df_1 df_2 \\ &= \frac{A_{eq}^2}{4\pi\beta_2\alpha_{eq}} \times \text{asinh}\left(\frac{\pi^2\beta_2 B_{WDM}^2}{4\alpha_{eq}}\right) \end{aligned} \qquad \text{Eq. (19)}$$

Replacing Eq. (17) and Eq. (18) in Eq. (19) we then have:

$$\begin{aligned} G_{NLI}(0) &\cong \frac{4}{27\pi}\gamma^2 G_{WDM}^3 \times \frac{(1 - e^{-2\alpha L})^3}{[1 - e^{-2\alpha L} - 2\alpha L e^{-2\alpha L}] \times \beta_2\alpha} \\ &\quad \times \text{asinh}\left(\frac{\pi^2 \times [1 - e^{-2\alpha L} - 2\alpha L e^{-2\alpha L}] \times \beta_2 B_{WDM}^2}{4\alpha \times (1 - e^{-2\alpha L})}\right) \end{aligned} \qquad \text{Eq. (20)}$$

It is obvious from Eq. (20) that for high loss condition where $e^{-2\alpha L} \ll 1$ Eq. (20) will converge to Eq. (8).

# Multi-Span Link

For a multi-span link, with identical spans and an amplifier at the end of each span that completely compensates for the span loss, the GN model formula is [2]:

$$G_{NLI}(f) = \frac{16}{27}\gamma^2 \int_{-\infty}^{+\infty} \int_{-\infty}^{+\infty} G_{WDM}(f_1) G_{WDM}(f_2) G_{WDM}(f_1 + f_2 - f)$$
$$\times \left| \frac{1 - e^{(-2\alpha + j\xi)L}}{2\alpha - j\xi} \times \frac{\sin(0.5 \times N_s \times L \times \xi)}{\sin(0.5 \times L \times \xi)} \right|^2 df_1 df_2 \quad \text{Eq. (21)}$$

Where $N_s$ is the number of spans in the link and $\xi$ was presented in Eq. (2). At $f = 0$, for an overall WDM rectangular spectrum as presented in Eq. (3) and changing the integration domain from lozenge shape to a square shape we have:

$$G_{NLI}(0) \cong$$
$$\frac{16}{27}\gamma^2 G_{WDM}^3 \int_{-\frac{B_{WDM}}{2}}^{+\frac{B_{WDM}}{2}} \int_{-\frac{B_{WDM}}{2}}^{+\frac{B_{WDM}}{2}} \left| \frac{1 - e^{(-2\alpha + j\xi)L}}{2\alpha - j\xi} \times \frac{\sin(0.5 \times N_s \times L \times \xi)}{\sin(0.5 \times L \times \xi)} \right|^2 df_1 df_2 \quad \text{Eq. (22)}$$

2-D integration in Eq. (22) does not have a closed-form solution. We define:

$$F_4(x) \triangleq \frac{1 - e^{(-2\alpha + x)L}}{2\alpha - x} \times \frac{e^{0.5 N_s L x} - e^{-0.5 N_s L x}}{e^{0.5 L x} - e^{-0.5 L x}} \quad \text{Eq. (23)}$$

where in Eq. (23):

$$x = j\xi = j4\pi^2 \beta_2 f_1 f_2 \quad \text{Eq. (23)}$$

The Taylor series of $F_4(x)$ around x=0 is written as:

$$F_4(x) \triangleq \left[ N_s \times \frac{1 - e^{-2\alpha L}}{2\alpha} \right]$$
$$+ \left\{ \frac{N_s \times [1 - e^{-2\alpha L} - 2\alpha L e^{-2\alpha L}]}{4\alpha^2} + \frac{N_s(N_s - 1)L \times [1 - e^{-2\alpha L}]}{4\alpha} \right\} x + \cdots \quad \text{Eq. (24)}$$

Considering Eq. (16) and inserting the two first Taylor terms for $F_4(x)$ and $F_3(x)$ we have:

$$\alpha_{eq} = \frac{\alpha \times (1 - e^{-2\alpha L})}{1 - e^{-2\alpha L} - 2\alpha L e^{-2\alpha L} + \alpha L (N_s - 1)(1 - e^{-2\alpha L})} \quad \text{Eq. (25)}$$

$$A_{eq} = \frac{N_s \times (1 - e^{-2\alpha L})^2}{1 - e^{-2\alpha L} - 2\alpha L e^{-2\alpha L} + \alpha L(N_s - 1)(1 - e^{-2\alpha L})} \quad \text{Eq. (26)}$$

The GN model can then be approximated as:

$$G_{NLI}(0) \cong \frac{16}{27}\gamma^2 G_{WDM}^3 \int_{-\frac{B_{WDM}}{2}}^{+\frac{B_{WDM}}{2}} \int_{-\frac{B_{WDM}}{2}}^{+\frac{B_{WDM}}{2}} \left| \frac{1 - e^{(-2\alpha + j\xi)L}}{2\alpha - j\xi} \times \frac{\sin(0.5 \times N_s \times L \times \xi)}{\sin(0.5 \times L \times \xi)} \right|^2 df_1 df_2$$

$$\cong \frac{16}{27}\gamma^2 G_{WDM}^3 \int_{-\frac{B_{WDM}}{2}}^{+\frac{B_{WDM}}{2}} \int_{-\frac{B_{WDM}}{2}}^{+\frac{B_{WDM}}{2}} \left| \frac{A_{eq}}{2\alpha_{eq} - j\xi} \right|^2 df_1 df_2$$

$$\cong \frac{16}{27}\gamma^2 A_{eq}^2 G_{WDM}^3 \int_{-\frac{B_{WDM}}{2}}^{+\frac{B_{WDM}}{2}} \int_{-\frac{B_{WDM}}{2}}^{+\frac{B_{WDM}}{2}} \left| \frac{1}{2\alpha_{eq} - j\xi} \right|^2 df_1 df_2 \quad \text{Eq. (27)}$$

Therefore replacing $\alpha$ with $\alpha_{eq}$ in Eq. (8) and scaling it with $A_{eq}^2$ we have:

$$G_{NLI}^{generalized}(0) \cong \frac{4}{27\pi} \times \frac{(1 - e^{-2\alpha L})^3}{[1 - e^{-2\alpha L} - 2\alpha L e^{-2\alpha L} + \alpha L(N_s - 1)(1 - e^{-2\alpha L})]} \times \frac{\gamma^2 N_s^2 G_{WDM}^3}{\alpha \beta_2}$$
$$\times \operatorname{asinh}\left( \frac{\pi^2 \beta_2 B_{WDM}^2}{4} \times \frac{[1 - e^{-2\alpha L} - 2\alpha L e^{-2\alpha L} + \alpha L(N_s - 1)(1 - e^{-2\alpha L})]}{\alpha \times (1 - e^{-2\alpha L})} \right) \quad \text{Eq. (28)}$$

## Validity Range of Eq. (28)

In the procedure of deriving Eq. (28), we tried to replace the function under integral in Eq. (22) with an approximate function which is analytically integrable. This similarity happens at x=0 because we wrote the Taylor series at x=0 and then imposed that the two 1$^{st}$ terms be equal.

We must note that the term $\left|\frac{\sin(0.5 \times N_s \times L \times \xi)}{\sin(0.5 \times L \times \xi)}\right|^2$ in Eq. (22) is a periodic function whose maximum value occurs at $\xi = 0$ and $\xi = \frac{2k\pi}{L}$ where k is an integer. Clearly, we tried to obtain an approximation of the integrand function in Eq. (22) near $\xi = 0$. For this approximation to be accurate, we need at least one of the two below conditions to hold:

1. The integration area only contains the maximum of $\left|\frac{\sin(0.5 \times N_s \times L \times \xi)}{\sin(0.5 \times L \times \xi)}\right|^2$ which is at $\xi = 0$ but none of the other maxima, that occur at $\xi = \frac{2k\pi}{L}$ for $k \neq 0$. This is equivalent to requiring that: $|\pi L \beta_2 B_{WDM}^2| < 2$. A quick calculation shows that this condition is not easily met.

2. The maxima at $\xi = \frac{2k\pi}{L}$ for $k \neq 0$ are much lower than one at $\xi = 0$, so that we can ignore their contribution in the integral: $\left|\frac{1-e^{-2\alpha L}}{2\alpha - j(\frac{2\pi}{L})}\right| \ll \left|\frac{1-e^{-2\alpha L}}{2\alpha}\right|$. This is equivalent to requiring that $\alpha L \ll \pi$. This condition is met over short spans for conventional loss (about 10 km) or if loss is much lower than conventional silica fiber.

## Incoherent Approximation for multi-span Links

One common approximation in calculating NLI using the GN model is considering the NLI of each span independent of that of the other spans. Accepting this approximation, which is called "incoherent approximation," the NLI of a multi-span link can be calculated approximately by calculating the NLI contribution due to each span independently and then simply adding all contributions at the end of link. Therefore, considering Eq. (20) as the contribution of a single span and accepting the incoherent NLI accumulation approximation for a multi-span link containing $N_s$ spans, with identical spans and an amplifier at the end of each span that completely compensates for the span loss, we can write:

$$G_{NLI}^{improved}(0) \cong \frac{4}{27\pi} \gamma^2 G_{WDM}^3 \times \frac{N_s \times (1 - e^{-2\alpha L})^3}{[1 - e^{-2\alpha L} - 2\alpha L e^{-2\alpha L}] \times \beta_2 \alpha} \times \mathrm{asinh}\left(\frac{\pi^2 \times [1 - e^{-2\alpha L} - 2\alpha L e^{-2\alpha L}] \times \beta_2 B_{WDM}^2}{4\alpha \times (1 - e^{-2\alpha L})}\right)$$

Eq. (29)

It can be shown that the above-mentioned approximation is a lower bound for the NLI assessment. Because NLI contributions of different spans can be added coherently or even partly coherent, the amount of NLI will always be equal or higher than a fully incoherent scenario (presented in Eq. (29)). Also, it is worth mentioning that when $|\pi L \beta_2 B_{WDM}^2| > 2$ the incoherent approximation is accurate. Therefore, considering the validity range for Eq. (28), mentioned in the previous section, we propose the general formula below, that holds without any limitations on span-length, number of spans, span loss, bandwidth of WDM comb and dispersion of the fiber as:

$$G_{NLI} \cong \max\left(G_{NLI}^{improved}(0), G_{NLI}^{generalized}(0)\right)$$

Eq. (30)

# Numerical Results

To validate the accuracy of Eq. (30), we integrated the exact GN model formula Eq. (21) numerically and compared the results of the NLI power spectral density at the center of the comb with Eq. (30). We considered a 10-span homogenous (all-identical spans) and transparent (loss is exactly compensated for at the end of each span) optical link. The fiber nonlinearity parameter is $\gamma = 1.2\ (W.Km)^{-1}$ and fiber dispersion is $\beta_2 = -21\ ps^2/km$. The WDM comb is considered as one seamless rectangle with 5 THz bandwidth and power spectral density (PSD) equal to $G_{WDM} = 1\ W/THz$. Then the attenuation parameter of the fiber is varied from $5 \times 10^{-4}$ to $0.3\ dB/km$. Different span lengths are addressed, ranging from 100m to 1000km. For each span length, the results of the GN model numerical integration are compared with the results of Eq. (30) for 1 and 10 spans in the link.

The series of plots below show remarkable accuracy, as the error always remains below 0.5 dB in all cases.

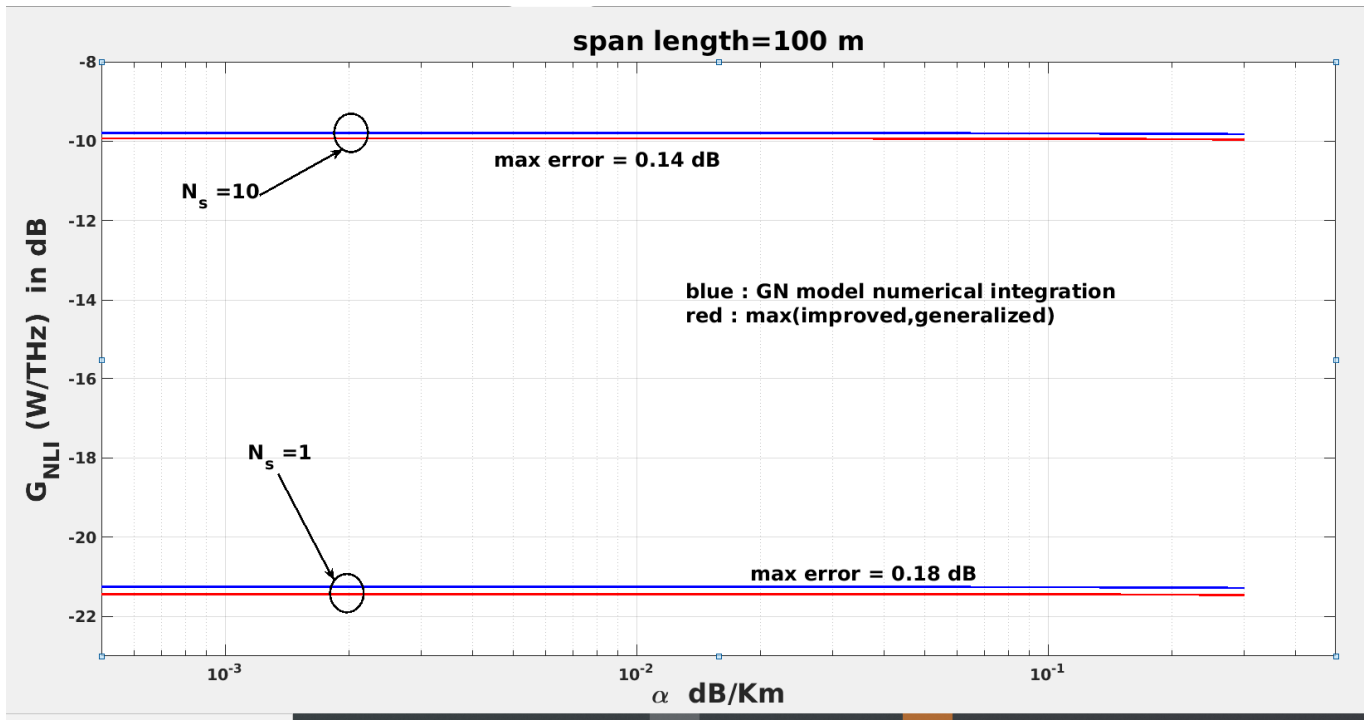

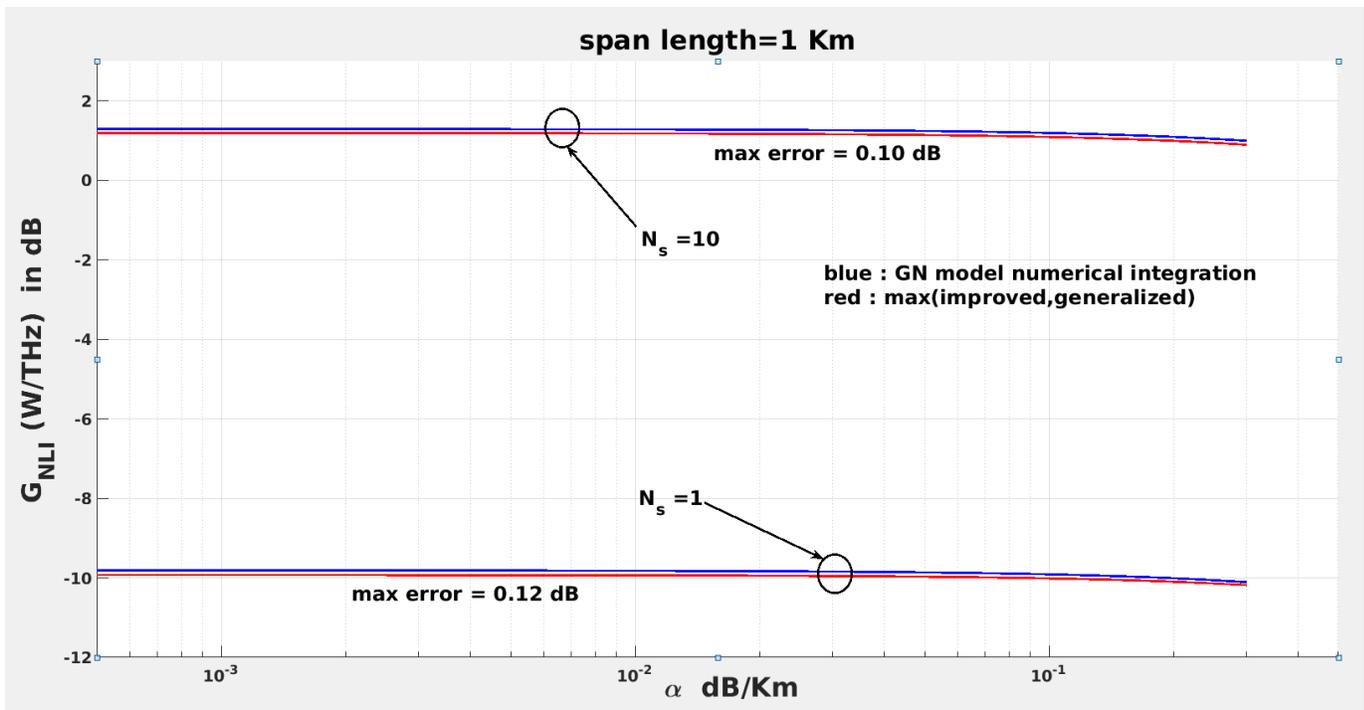
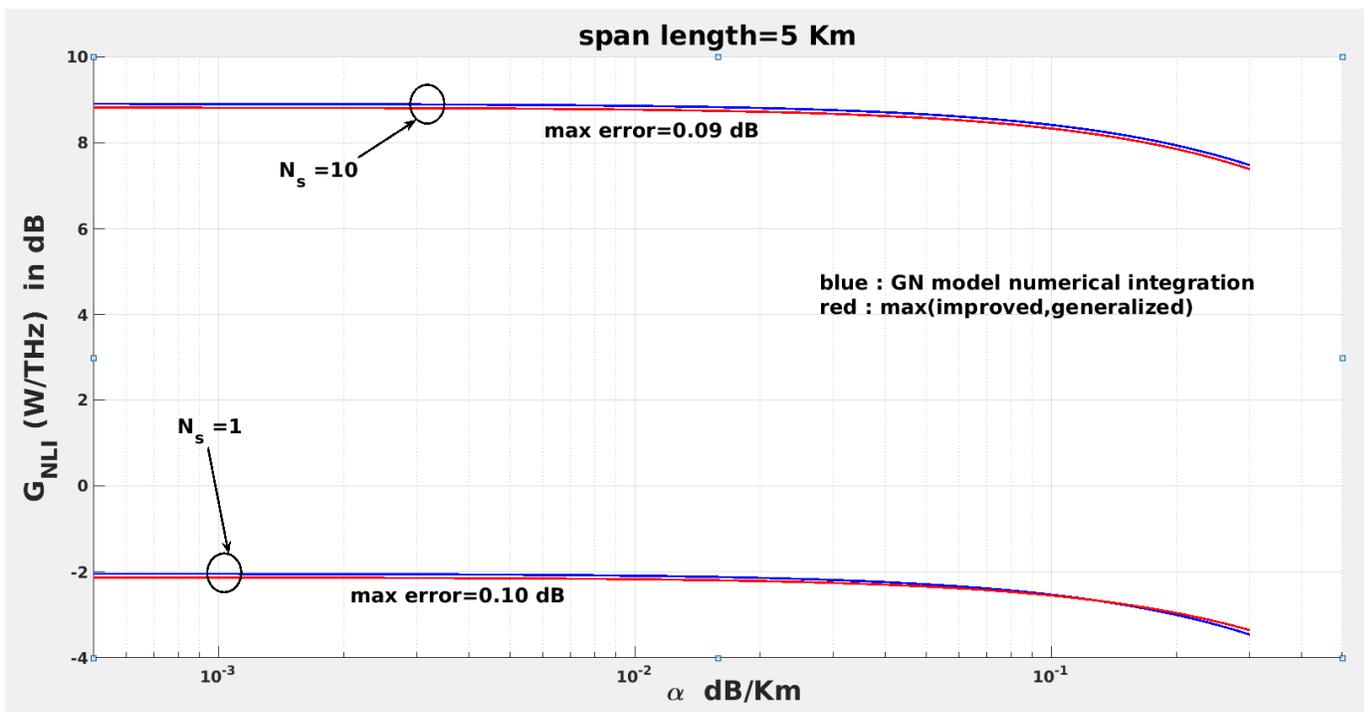

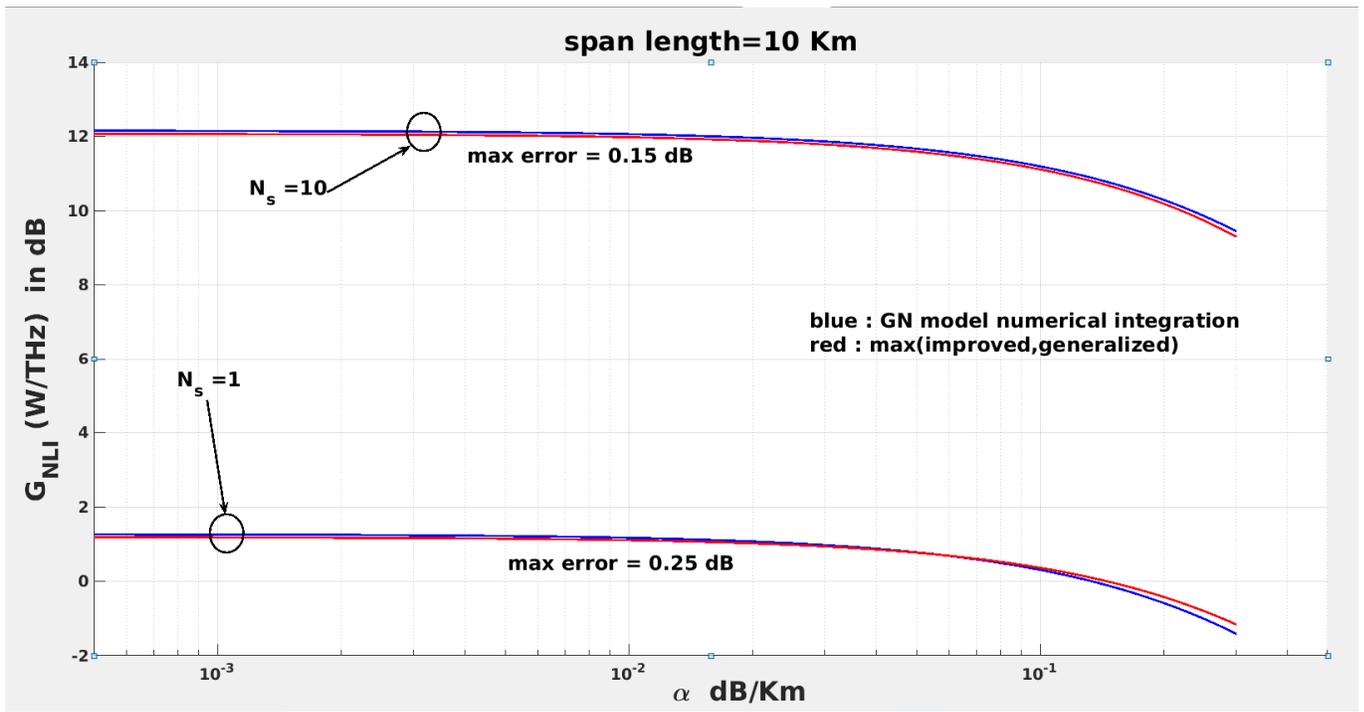

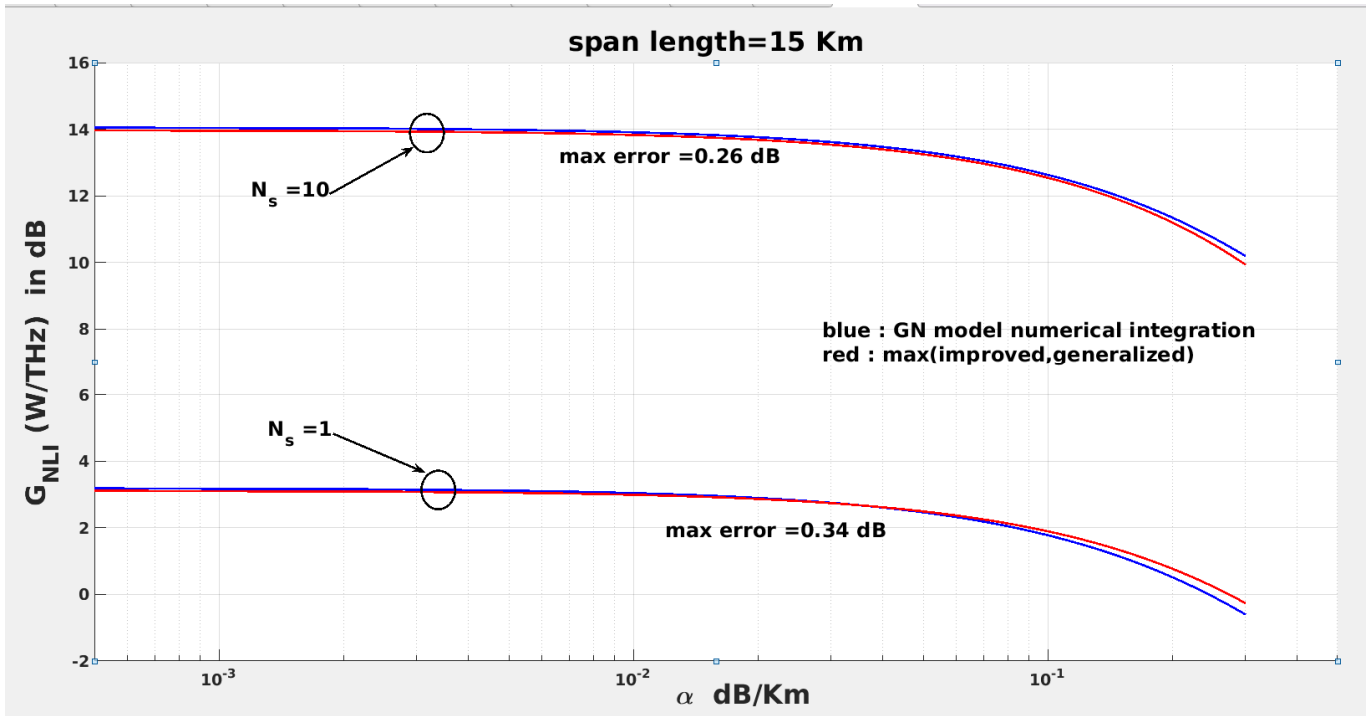

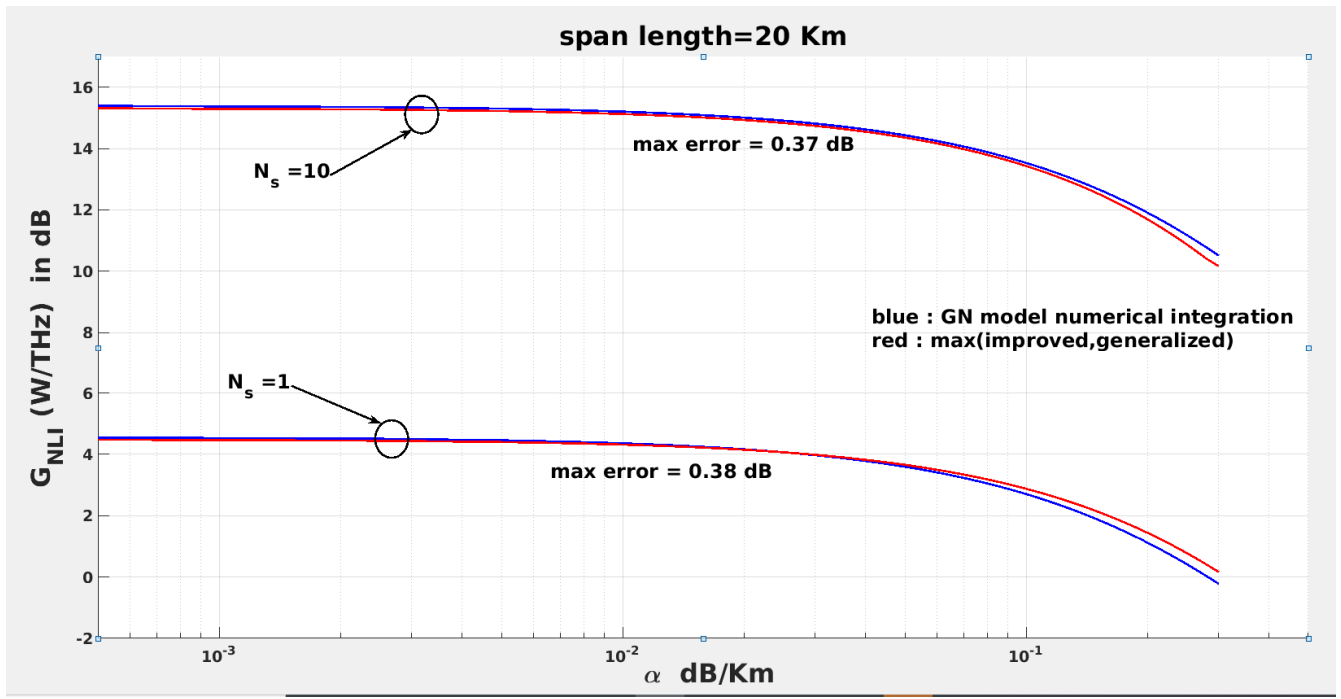
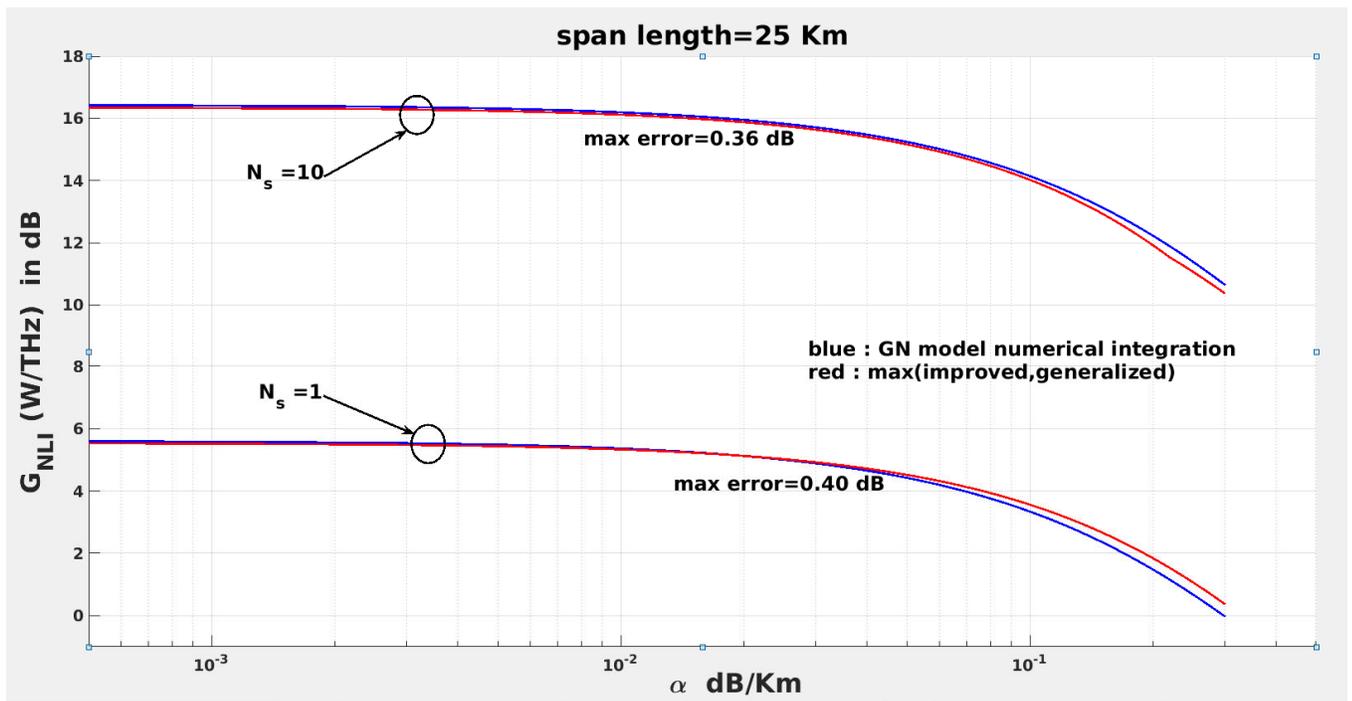

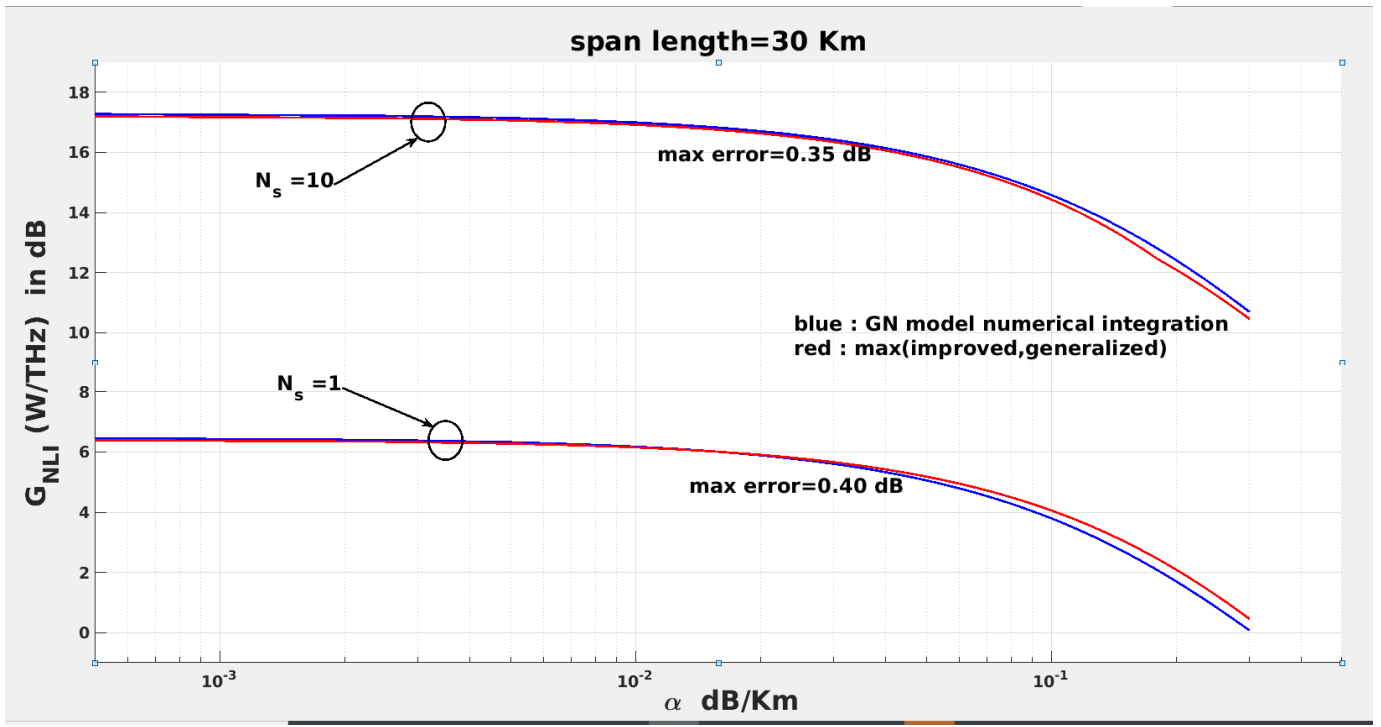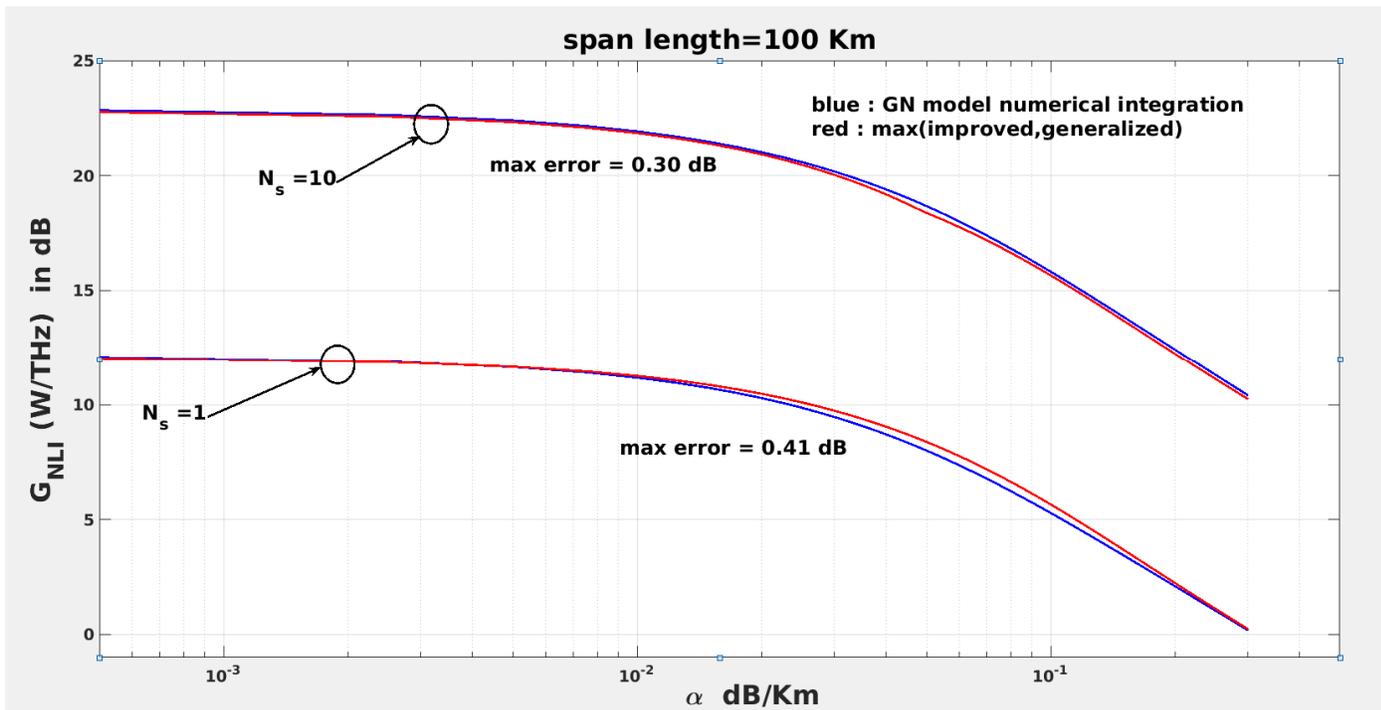

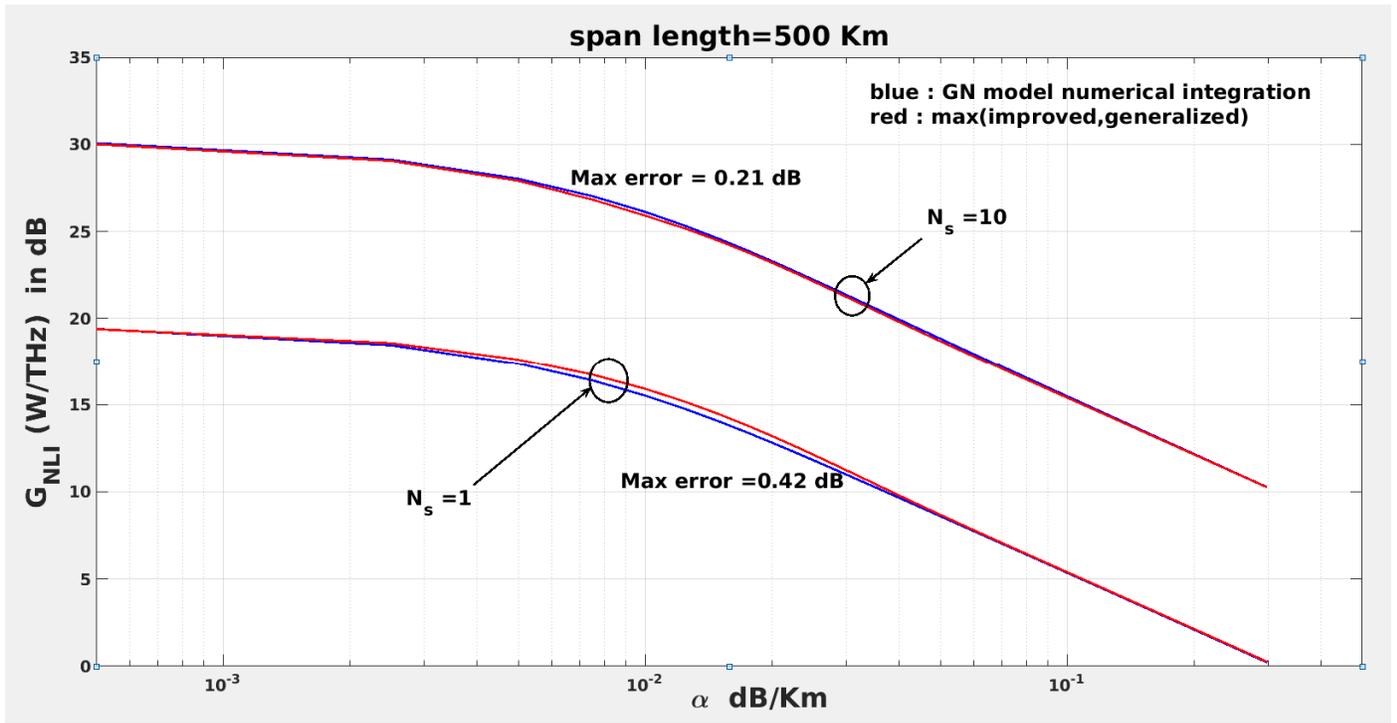

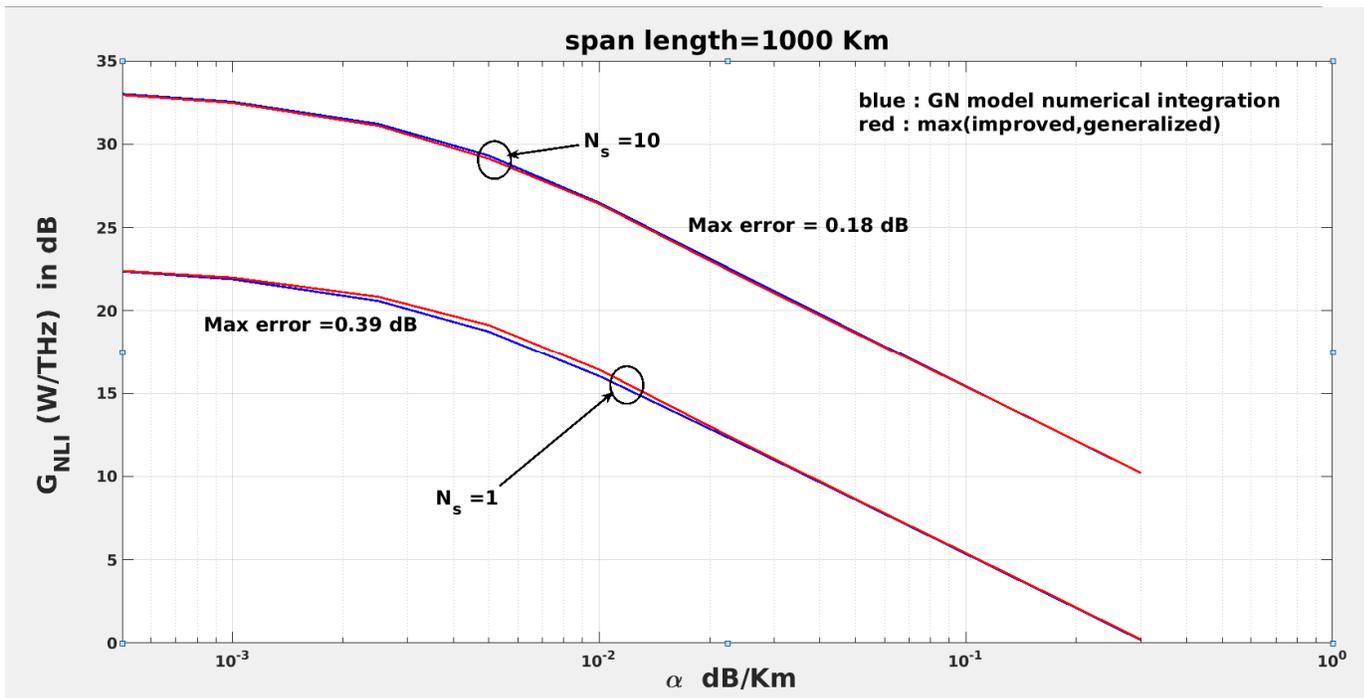

# Conclusion

In this paper we have successfully carried out the derivation of new closed-form formulas that approximate the GN-model integral in fiber and span loss regimes that were not addressed by previous closed-form formulas. Notably, the low-loss and low-span-loss regimes, that are needed to discuss future hollow-core fiber scenarios as well as to solve current modeling problems such as the presence of lumped loss within conventional fiber spans.